\documentclass
[preprint,byrevtex,nofootinbib,10pt,twocolumn,balancelastpage,titlepage,bibnotes]{revtex4}%
\usepackage{amsfonts}
\usepackage{amsmath}
\usepackage{amssymb}
\usepackage{graphicx}%
\providecommand{\U}[1]{\protect\rule{.1in}{.1in}}

\ifx\pdfoutput\relax\let\pdfoutput=\undefined\fi
\newcount\msipdfoutput
\ifx\pdfoutput\undefined\else
\ifcase\pdfoutput\else
\ifx\paperwidth\undefined\else
\ifdim\paperheight=0pt\relax\else\pdfpageheight\paperheight\fi
\ifdim\paperwidth=0pt\relax\else\pdfpagewidth\paperwidth\fi
\fi\fi\fi
\begin{document}
\preprint{ }
\preprint{UATP/2203}
\title{Maxwell's Conjecture of the Demon creating a Temperature Difference is False}
\author{P.D. Gujrati,}
\affiliation{$^{1}$Department of Physics, $^{2}$School of Polymer Science and Polymer
Engineering, The University of Akron, Akron, OH 44325}
\email{pdg@uakron.edu}

\begin{abstract}
We argue that Maxwell's demon is incapable of creating a nonzero temperature
difference. Hence, it does not destroy equilibrium and the second law is never
at risk, contrary to the claim by Maxwell and accepted by many. It is
therefore remarkable that despite this, the demon paradox has been a valuable
source of new ideas. We use two independent arguments, one using classical
equilibrium thermodynamics by extending Brillouin's approach, and the other
one using equilibrium statistical mechanics and the central limit theorem.

\end{abstract}
\date{March 25, 2022}
\maketitle

Maxwell's demon, which has been puzzling scientists since 1867, stands between
two neighboring chambers $\Sigma_{1}$ and $\Sigma_{2}$ (having fixed and
identical volumes) sharing a thermally insulating and impenetrable wall with a
hole \cite{Knott,Maxwell,Smoluchowski,Feynman,Maruyama,Rex}, and having an
ideal gas containing $N$ particles. The demon D opens or closes the small hole
at will to select faster particles to go from $\Sigma_{2}$\ into $\Sigma_{1}$
and slower particles from $\Sigma_{1}$\ into $\Sigma_{2}$; Maxwell does not
describe how D knows their energies. The chambers and D form an isolated
system $\Sigma$ of volume $V$. The wall, the hole, and D act like an inert
piston in a cylinder of gas is treated \cite{Prigogine}. Thus, we will focus
only the ideal gas, whose entropy we denote by $S$. Maxwell considered
$\Sigma$ to be initially in equilibrium (EQ) having a temperature $T$ before D
intervenes. He \emph{conjectured} as self-evident that after a while
($\Delta\tau>0$), D creates a nonzero temperature difference $\Delta
T=T_{1}-T_{2}>0$ between the temperatures $T_{1}$ and $T_{2}$ of $\Sigma_{1}%
$\ and $\Sigma_{2}$, respectively, without any expenditure of work
\cite{Maxwell}. As the gas is no more in EQ, its entropy $S$ must decrease, a
violation of the second law (the \emph{demon paradox}); for an open system,
there may be no violation. The paradox has generated a tremendous amount of
debate and some confusion among the best minds of our time since its inception
\cite{Rex}. The endeavor has been a constant source of major conceptual
advances and some challenges in theoretical physics
\cite{Brillouin,Landauer,Szilard,Bennett,Plenio} and in the philosophy of
science \cite{Norton,Bennett0} that has resulted in the demon problem to
undergone many modifications including\ those requiring an open $\Sigma$
\cite{Brillouin,Landauer,Szilard}.

The main source of confusion in these developments has been the concept of any
work done by the demon as it performs "measurements" of energies, and has
required the concepts of feedback mechanism, information (including mutual
information) entropy, Landauer's principle, minimum dissipation, erasure, etc.
The most common response is that D somehow manages to create enough entropy to
salvage the second law \cite{Brillouin,Landauer,Szilard,Bennett,Plenio} but it
is just as common for not everyone to agree to the actual manner in which this
happens \cite{Kish,Shenker,Norton,Norton1,Bennett0,Porod,Toffoli,Benioff}.
Maxwell's conjecture $\Delta T>0$ seems very natural and almost self-evident
because of the separation of more energetic (fast, $\epsilon_{\text{f}}>3T/2$)
and less energetic (slow, $\epsilon_{\text{s}}<3T/2$) particles, and is
universally accepted without ever being questioned; here $\epsilon_{\text{mp}%
}=3T/2$ is the most probable energy per particle. This is surprising as
equilibrium thermodynamic and statistical mechanics tells us that for a
macroscopic system, only the most probable state matters; the improbable
states are almost irrelevant. In view of this, we have decided to carefully
investigate the conjecture. A negative answer will make the above
modifications, though interesting and mentally challenging in their own
rights, not physically or conceptually relevant for solving the demon paradox.
Our interest is not to continue the above controversies as they do not pertain
to the verification of Maxwell's conjecture in an isolated system.\ We will
also ask if there is a need for any measurement and erasure, and if the loss
of equilibrium is mere fluctuation prior to reset. \ \ %

\begin{figure}
[ptb]
\begin{center}
\includegraphics[
height=2.8945in,
width=3.998in
]%
{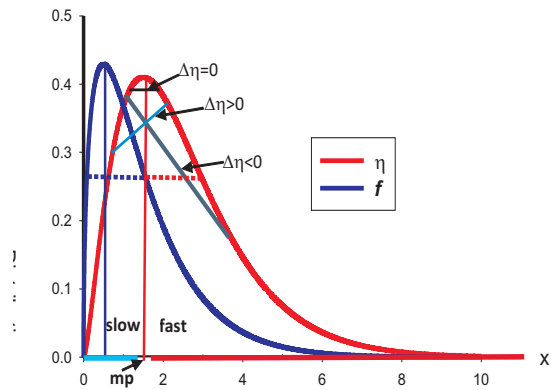}%
\caption{Probability density functions $f(x)$ (blue curve) and $\eta(x)$ (red
curve) in terms of adimensional energy $x=\epsilon/T$, and the distribution of
slow (light blue solid line) and fast (dark red solid line) along $x$-axis
with a gap mp between them at the solid red vertical line representing most
probable energy $x_{\text{mp}}$, which do not participate in any flow across
the hole. Three lines touching the red curve give energy $\Delta\eta$
transported across the open hole of any sign. }%
\label{Fig-Dist}%
\end{center}
\end{figure}
We will closely follow Brillouin \cite{Brillouin}, who seems to be the first
to treat particle flows carefully assuming $\Delta T>0$, but the main interest
was to determine the entropy generated by observation through light. He does
not account for how often particles pass through the open door and how much
average energies they transport during $\Delta\tau$. This is unfortunate as
slow particles will go through more often than the fast particles so it is
possible for them to transport more energy than the fast ones; see Eq.
(\ref{Brillouin0}) below. This issue has never been discussed for the demon
paradox so far. This is where we deviate from Brillouin.

We show by two independent methods that D cannot create a nonzero $\Delta
T$\ and destroy EQ. Thus, D remains in EQ with the gas and there is no need
for $\emph{resetting}$ D. The two approaches are (i) classical equilibrium
thermodynamics with no fluctuations and (ii) equilibrium statistical mechanics
involving fluctuations. We say nothing about an interacting $\Sigma$, which we
will discuss elsewhere.

The average number of particles $dN_{x}$ per unit volume with the
(adimensional) energy $x=\epsilon/T$ due to translation motion in the interval
$dx$\ is given by \cite{Reif,Callen,Landau} $dN_{x}=(2N/V)\sqrt{x/\pi}%
e^{-x}dx$ with $%
{\textstyle\int\nolimits_{0}^{\infty}}
dN_{x}=N/V$. The average energy carried by these particles is $xdN_{x}$. It
will be useful to suppress unnecessary constant terms and introduce the
following two unnormalized probability distribution functions (pdf's)%
\begin{equation}
f(x)=\sqrt{x}e^{-x},\eta(x)=xf(x)=x^{3/2}e^{-x}. \label{Distributions0}%
\end{equation}
Here, $f(x)$ is the pdf for (the random variable) $x$, and $\eta(x)$ for the
energy carried by these particles. They are shown by the blue and red curves
in Fig. \ref{Fig-Dist}. The areas under curves are $\Gamma(3/2)$ and
$\Gamma(5/2)$, respectively, with the ratio $\Gamma(5/2)/\Gamma(3/2)$ giving
the average energy $\overline{x}=3/2$. Each function has a single maximum:
$f_{\text{max}}\approx0.4289$ at $x=x_{\text{m}}=1/2$ shown by the location of
the blue vertical line, and $\eta_{\text{max}}\approx0.4099$ at $x=\overline
{x}$ ($\overline{\epsilon}$ $=3T/2$) shown by the location of the red vertical
line; here, $x_{\text{m}}$ is the most probable value, the mode of $f(x)$ and
is different from the mean $\overline{x}\doteq\epsilon_{\text{mp}}/T$. [It
should be remarked that $f(x)$ can also be interpreted as the average energy
per degree of freedom of the particle in the interval $dx$\ so that the
location of the blue line denotes the average energy $\bar{x}_{\text{df}}=1/2$
per degree of freedom over $\left(  0,\infty\right)  $.]

We now follow slow ($x_{\text{s}}$) and fast ($x_{\text{f}}$) particles. We
assume that D has very sharp faculty but not so sharp that he can determine
the particle energy with infinite precision. We introduce a very small but
nonzero positive quantity $\delta\sim10^{-10}$-$10^{-11}$ to be the limit of
D's precision \cite{Note0} so that all particles $x_{\text{mp}}$ lie in a
window $\delta\overline{x}\doteq\overline{x}-\delta\leq x\leq\overline
{x}+\delta$. Particles $x_{\text{s}}$ with $x<\overline{x}-\delta$ and
$x_{\text{f}}$ with $x>\overline{x}+\delta$ occupy, respectively, the solid
light blue and the solid dark red portions of the $x$-axis in Fig.
\ref{Fig-Dist}. D only manipulates these particles but \emph{not} the
$x_{\text{mp}}$ in the window $\delta\overline{x}$ on the $x$-axis. The dotted
blue-red horizontal blue line at the height $f(\overline{x})\approx0.2733$
cuts the blue curve at $x_{1}\approx0.0895$ and $x=$ $\overline{x}$, and the
red curve at $x_{2}\approx0.6492$ and $x_{3}\approx2.8835$. We see that for
$x\geq x_{1}$, $f(x_{\text{s}})\geq f(x_{\text{f}})$. Thus, slow particles
appear more often than the fast particles for $x\geq x_{1}$. From $\eta
(x)$\ over ($x_{2},x_{3}$), we can easily conclude that $\Delta\eta\doteq
\eta(x_{\text{f}})-\eta(x_{\text{s}})$ can have any sign, a fact that will be
important in thermodynamic consideration.

\textbf{Thermodynamics: }For simplicity, we will use the term "body" and
denoted by $\Sigma_{\text{b}}$ to refer to any one of $\Sigma,\Sigma_{1}$, and
$\Sigma_{2}$ as their thermodynamic discussion is very similar. The body
$\Sigma_{\text{b}}$ is described by the observables $\left(  E_{\text{b}%
},N_{\text{b}}\right)  $; its volume is kept fixed. As $\Sigma_{1}$ and
$\Sigma_{2}$ together require four observables, we also need the same number
to specify $\Sigma$. Using $E_{1}$\ and $E_{2}$, we introduce combinations
$E=E_{1}+E_{2}$ and $E^{\prime}=E_{1}-E_{2}$; similarly, we introduce
$N=N_{1}+N_{2}$ and $N^{\prime}=N_{1}-N_{2}$. They now uniquely specify
$\Sigma$. As $E$ and $N$ are fixed, only $E^{\prime}=2E_{1}-E$ and $N^{\prime
}=2N_{1}-N$\ will change. Let us consider the changes $dE^{\prime}=2dE_{1}$
and $dN^{\prime}=2dN_{1}$, which are caused by transfer of fast and slow
particles so that%
\begin{equation}
dE_{1}=dE_{\text{f}}-dE_{\text{s}},dN_{1}=dN_{\text{f}}-dN_{\text{s}}.
\label{Energy Transfer0}%
\end{equation}
It should be noted that $dE_{\text{f}}$ and $dE_{\text{s}}$ or $dN_{\text{f}}$
and $dN_{\text{s}}$ cannot represent two independent variations as their
differences are single variations $dE_{1}$ and $dN_{1}$, respectively. Let us
consider their significance. From $S(E,N,E^{\prime},N^{\prime})=S_{1}%
(E_{1},N_{1})+S_{2}(E_{2},N_{2})$, we have
\begin{equation}
dS=(\beta_{1}-\beta_{2})dE_{1}+(\mu_{2}\beta_{2}-\mu_{1}\beta_{1})dN_{1}
\label{Entropy Transfer0}%
\end{equation}
in $\Sigma$, where $\beta_{1}=1/T_{1}=\partial S_{1}/\partial E_{1}$ and
$\beta_{2}=1/T_{2}=\partial S_{2}/\partial E_{2}$, and $\beta_{1}\mu
_{1}=-\partial S_{1}/\partial N_{1}$ and $\beta_{2}\mu_{2}=-\partial
S_{2}/\partial N_{2}$\ of $\Sigma_{1}$ and $\Sigma_{2}$. For the second law to
be valid, each term on the right must be nonnegative; otherwise not. In
equilibrium, they both vanish identically.

Brillouin \cite{Brillouin} also uses $\overline{x}$ as the cutoff for slow and
fast particles as we have done. However, he merely follows Maxwell's
conjecture that after a certain time $\Delta\tau$, D creates nonzero $\Delta
T<<T$. He then proceeds with one fast particle moving from $\Sigma_{2}$ into
$\Sigma_{1}$, and one slow particle from $\Sigma_{1}$ into $\Sigma_{2}$, and
finds $\Delta Q=3(\epsilon_{1}+\epsilon_{2})/2>0$ without accounting for their
probabilities; here, we are using the notation of Brillouin. As fast particles
have smaller probabilities than the slow particles, see $f(x)$\ in Fig.
\ref{Fig-Dist}, it is possible that D selects two slow particles for a fast
particle so that the net energy transfer is
\begin{equation}
\Delta Q=3(-1+\epsilon_{1}+2\epsilon_{2})/2<0, \label{Brillouin0}%
\end{equation}
which results in $\Delta S_{i}>0$. This is different from $\Delta Q>0$ above
and resulting in $\Delta S_{i}<0$ \cite[Eqs. (13) and (14)]{Brillouin}.
However, it is not correct to interpret $\Delta S_{i}>0$ \ as supporting the
second law and $\Delta S_{i}<0$ as violating the second law as $\Delta S_{i}$
is evaluated using a few (one or two) particles.

We see from the above discussion that it is not just the energies but also
probabilities of the particles play important roles in determining the energy
transfer. Thus, we need to consider $\Delta\eta(x)$, which is the analog of
the quantity $\Delta Q$ of Brillouin for two particles. Consider the solid
black horizontal line near the peak of $\eta(x)$ that meets the red curve at a
slow and a fast particles having the same energy to give $\Delta\eta=0$. By
taking the black line at different angles will give us $\Delta\eta$ (or
$\Delta Q$) of both signs. Thus, there is no one particular sign of
$\Delta\eta$. This casts doubts on the claim by Maxwell and supported by
Brillouin's calculation that $\Sigma_{1}$ has a higher temperature than
$\Sigma_{2}$ after many particle transfers during $\Delta\tau$. Indeed, it is
just as possible to have the opposite situation as seen from Eq.
(\ref{Brillouin0}). In this case, the flow of two slow particles into
$\Sigma_{2}$\ raises its energy over $\Sigma_{1}$by more than the increase in
the energy of $\Sigma_{1}$ by one fast particle over the same duration of time.

Indeed, it is quite possible that during $\Delta\tau$, the fluctuating signs
of $\Delta\eta$ yield a zero average so there is no energy difference between
$\Sigma_{1}$ into $\Sigma_{2}$. We first recognize that the concept of
temperature, heat flow, entropy, entropy generations, etc. are macroscopic
concepts so D must let a very large number of particles through over
$\Delta\tau$ and average over them to give%
\begin{equation}
dE_{1}=a_{1}dT_{1}+a_{2}d\mu_{1},dN_{1}=b_{1}dT_{1}+b_{2}d\mu_{1},
\label{Energy Transfer1}%
\end{equation}
by treating $E_{1}$ and $N_{1}$\ equivalently as functions of $T_{1}$ and
$\mu_{1}$ of $\Sigma_{1}$; the coefficients are also functions of $T_{1}$ and
$\mu_{1}$. For $\Sigma_{2}$, we have similar relations with the same EQ
coefficients but with $dT_{2}=-dT_{1}$ and $d\mu_{2}=-d\mu_{1}$.

As the presence of $x_{\text{mp}}$ is never affected, $dE_{1}$ and $dN_{1}$
must be expressed as in Eq. (\ref{Energy Transfer0}). Using Eq.
(\ref{Entropy Transfer0}), we finally obtain
\begin{equation}
dS=\left[  \Delta\beta a_{1}-\Delta(\mu\beta)b_{1}\right]  dT_{1}+\left[
\Delta\beta a_{2}-\Delta(\mu\beta)b_{2}\right]  d\mu_{1},
\label{Entropy Transfer1}%
\end{equation}
where $\Delta\beta=\beta_{1}-\beta_{2}$ and $\Delta(\mu\beta)=\mu_{1}\beta
_{1}-\mu_{2}\beta_{2}$. We now apply this equation to the initial EQ state of
$\Sigma_{\text{b}}$, just when D begins to sort out particles. As $\Delta
\beta=0$ and $\Delta(\mu\beta)=0$ in the EQ state, we see that $dS_{\text{b}%
}=0$ so D does not affect $S,S_{1}$, and $S_{2}$. This means that the EQ is
not destroyed and $T_{\text{b}}=T,\mu_{\text{b}}=\mu$ during $\Delta\tau$. As
$dT_{1}=0$ and $d\mu_{1}=0$, we see that even $dE_{1}=0$ and $dN_{1}=0$ so
\begin{equation}
dE_{\text{f}}=dE_{\text{s}},dN_{\text{f}}=dN_{\text{s}} \label{dE-dN}%
\end{equation}
over $\Delta\tau$. It is now clear that the situation does not change at all
no matter how long D operates the hole. We have thus \emph{falsified}
Maxwell's conjecture. We emphasize that the above argument does not require
using the second law as we did not exploit the nonnegativity of the two terms
in Eq. (\ref{Entropy Transfer0}).

\textbf{Statistical Mechanics:}\ We now give the statistical mechanical
argument, which is independent of the above thermodynamic argument in that we
consider fluctuations that were neglected above. As D needs to average over
many particles, $f(x)$ and $\eta(x)$ of single particles are not useful. We
need to consider the entire gas, which requires considering the sum $X=%
{\textstyle\sum\nolimits_{i=1}^{N}}
x_{i}$ of the independent and identically distributed (the same\textit{\ }mean
$\overline{x}=3/2$ and the standard deviation $\sigma=\sqrt{3/2}$)\ random
variables $x_{i}$ of the $i$th\ particle. It follows from the central limit
theorem ($N>>1$) \cite{Reif} that $x\doteq X/N$, which is relevant for single
particles, has a \emph{normal distribution }$\mathcal{N}(\overline{x}%
,\sigma/\sqrt{N})$ with mean $\overline{x}$, where it also has its peak, and
the standard deviation $\sigma/\sqrt{N}$. Thus, $f(x)$ will approach the
central limit $\bar{f}(x)$ of a normal distribution:%
\begin{equation}
\bar{f}(x)=\frac{\sqrt{\pi}}{2}\mathcal{N}(\overline{x},\frac{\sigma}{\sqrt
{N}})=\sqrt{\frac{N}{8}}\frac{1}{\sigma}\exp[-\frac{N}{2}\left(
\frac{x-\overline{x}}{\sigma}\right)  ^{2}]; \label{Central Distribution0}%
\end{equation}
the prefactor $A_{f}=\sqrt{\pi}/2$ is inserted to ensure $\bar{f}(x)$ has the
same area as $f(x)$. Similarly, $\eta(x)$ will approach the central limit
$\bar{\eta}(x)=x$ $\bar{f}(x)$ with its peak at $\overline{x}$ as expected.
Oncoming particles towards the hole during $\Delta\tau$ are any of $N$
particles on average, so their pdf is $\bar{f}(x)$ and not $f(x)$. The maximum
of $\bar{f}(x)$ gives the most probable state for the thermodynamic
consideration, which as discussed below is very sharp with relative
fluctuations $\simeq1/\sqrt{N}$ that are almost vanishingly small due to the
presence of $N$ in the exponent and are neglected in thermodynamic
consideration above \cite[see discussion on pages 5 and 28]{Landau}. As is
well known, the mode, median and the mean for $\bar{f}(x)$ are all equal to
$\overline{x}$. A gas particle in equilibrium thermodynamics is governed not
by the broad maximum of $f(x)$ in Fig. \ref{Fig-Dist}, but by the sharp
maximum of $\bar{f}(x)$. This is consistent with Boltzmann's observation that
the average energy is determined by the most probable energy $3T/2$. The
entropy $S(\overline{E})$\ of the gas is also a function of the most probable
energy $\overline{E}=3NT/2$ with the inverse temperature given by the
thermodynamic derivative $1/T=\partial S(\overline{E})/\partial\overline{E}$.
Thus, even $S(\overline{E})$\ is determined by the peak at $\overline{x}$.

The discussion of $\bar{f}(x)$ above is very general and applies to any
$\Sigma_{\text{b}}$. Thus, we have three functions $\bar{f}_{\text{b}}(x)=($
$\bar{f}(x),\bar{f}_{1}(x),\bar{f}_{2}(x))$. Let the hole initially be open so
that all bodies have the same temperature. Consequently, all functions have
their peaks at $\overline{x}$. Each body has all kinds of
particles:\ $x_{\text{mp}},x_{\text{s}}$, and $x_{\text{f}}$. For any nonzero
$\delta$, we have ($\alpha=\left(  \delta/\sigma\right)  ^{2}/2>0$)%
\begin{subequations}
\begin{equation}
\bar{f}_{\text{b}}(x_{\text{s}})\text{ or }\bar{f}_{\text{b}}(x_{\text{f}%
})\lesssim\sqrt{N}e^{-N\alpha}\approx0. \label{CentralDistribution1}%
\end{equation}
Let us consider the case $N=10^{24}$ and $a=10^{-22}$ so that
\end{subequations}
\begin{equation}
\sqrt{N}e^{-N\alpha}=10^{12}e^{-100}\approx3.758\times10^{-32},
\label{CentralDistribution2}%
\end{equation}
a fantastically small number to justify the above approximation. Thus,
$x_{\text{s}}$ and $x_{\text{f}}$ particles have extremely low probabilities
and make no difference in determining the temperature, which is determined by
$x_{\text{mp}}$ alone.

The averaging over many particles is done using $\bar{f}_{\text{b}}(x)$ as
noted above. As a rule, D never allows $x_{\text{mp}}$ to be exchanged. This
means that the peak in $\bar{f}_{\text{b}}(x)$ remains in the same place at
$\overline{x}$ even after D begins to manipulate particle flows. Indeed, the
next to the most probable particles $x_{\text{mp}}$ are fast particles with
energies in the range $(\overline{x}+\delta<x_{\text{f}}^{\prime}<\overline
{x}+2\delta)$ and slow particles with energies in the range $(\overline
{x}-2\delta<x_{\text{f}}^{\prime}<\overline{x}-\delta)$, respectively. As we
see from Eqs. (\ref{CentralDistribution1}-\ref{CentralDistribution2}), the
pdf's for $x_{\text{s}}^{\prime}$ or $x_{\text{f}}^{\prime}$, although
themselves equal,$\ $are relative to that of $x_{\text{mp}}$ given by the
ratio $\bar{f}_{\text{b}}(x_{\text{s}}^{\prime}$ or $x_{\text{f}}^{\prime
})/\bar{f}_{\text{b}}(x_{\text{mp}})\approx e^{-N\alpha}\approx10^{-44}$. This
means that D will observe about $10^{44}$ most probable particles before
opening the hole to let a $x_{\text{s}}^{\prime}$ or $x_{\text{f}}^{\prime}$
particle through. As the temperature and chemical potentials of $\Sigma
_{\text{b}}$ are \emph{almost surely} determined by $x_{\text{mp}}$, and as D
has not affected $x_{\text{mp}}$ and the peak $\bar{f}_{\text{b}}%
(x_{\text{mp}})$, they are almost surely given by $T$ and $\mu$. In other
words, any rare transfer of slow and fast particles cannot affect the
temperature and chemical potential in $\Sigma_{\text{b}}$ so $E_{\text{b}}$
and $N_{\text{b}}$\ are also not affected. This immediately \emph{falsifies}
Maxwell's conjecture that seems to be universally accepted as was also
concluded thermodynamically in Eq. (\ref{dE-dN}). The statistical argument now
clarifies the significance of Eq. (\ref{dE-dN}) by rare events so it is
independent of the thermodynamic argument, and provides another support for
the falsification of Maxwell's conjecture for the simple reason that the
thermodynamic argument does not depend on the choice of $\delta$. As the
entropy $S_{\text{b}}$ of $\Sigma_{\text{b}}$ is also determined by the most
probable distribution, it has the same value as before D begins to operate.
This is again in accordance with the previous conclusion that $\Delta S=0$. We
have also arrived at the same conclusion earlier \cite{Gujrati-NEQ-MD} using a
nonequilibrium approach.

By focussing on $x_{\text{s}}$ and $x_{\text{f}}$ particles, Maxwell had hoped
to create energy imbalance between the chambers. But to suggest this
improbable energy imbalance results in a temperature imbalance does not work
as the temperature is a macroscopic concept defined more precisely by the
derivative $\left(  \partial E/\partial S\right)  _{\text{b}}$ or by $\bar
{f}_{\text{b}}(x_{\text{mp}})$. It is well known from the fluctuation theory
\cite{Landau} that fluctuations in $S_{\text{b}}$ such as due to $x_{\text{s}%
}^{\prime}$ or $x_{\text{f}}^{\prime}$ particle transfer occur at the same
temperature $T$ of $\Sigma_{\text{b}}$. The gas in each chamber even after
including these fluctuations will have the original $T$ describing $\bar
{f}_{\text{b}}(x_{\text{mp}})$. Thus, no perpetual motion machine of the
second kind can be formed by using equilibrium fluctuations in a system. As
$\Delta S=0$, there is no need to bring in the concept of information as there
is no second law violation to resolve. Whether we need it for Szilard's engine
is a separate issue since there we deal with a system in contact with a heat
bath \cite{Szilard,Bennett}. Our discussion here is for an isolated system for
which entropy reduction is a fundamental problem as Maxwell has observed. We
discuss the engine in a separate publication \cite{Gujrati-Szilard}.

As $\bar{f}_{\text{b}}(x)$ is symmetric about $x_{\text{mp}}$, $x_{\text{s}}$
and $x_{\text{f}}$ have the same probability distribution. Only if we start
with the two chambers at different temperatures will be obtain $dE_{1}\neq0$,
see Eq. (\ref{Energy Transfer1}), as was also discussed by Feynman
\cite{Feynman}. But this does not happen with the Maxwell's demon, where we
start with an equilibrium state. In our discussion, we have found no need to
worry about the mechanism used by D to observe particles and of intelligence
as was the case with Brillouin \cite{Brillouin}. As the initial equilibrium of
$\Sigma$ is not destroyed, D always remains in EQ with the two chambers so its
entropy cannot change, which justifies treating it as thermodynamically inert.
As the operations performed by D does not alter its initial state, there is no
\emph{resetting} required.

\end{document}